\documentclass[11pt, a4paper]{iopart}
\usepackage[utf8]{inputenc}

\usepackage{etoolbox}
\usepackage{graphicx}
\graphicspath{{.}{./Figures/}}

\makeatletter
\long\def\@makefntext#1{\parindent 1em\noindent 
 \makebox[1em][l]{\footnotesize\rm$\m@th{{}^\arabic{footnote}}$}%
 \footnotesize\rm #1}
\def\@makefnmark{\hbox{${}^{\arabic{footnote}}\m@th$}}
\def\@thefnmark{\arabic{footnote}}
\makeatother
\usepackage{fullpage}
\usepackage[linktocpage=true, colorlinks=true, linkcolor=blue, urlcolor=blue, citecolor=blue]{hyperref}
\usepackage{paralist}
\usepackage{xcolor}
\usepackage{soul}

\definecolor{darkbrown}{rgb}{0.4, 0.26, 0.13}
\definecolor{ao}{rgb}{0.0, 0.5, 0.0}
\definecolor{bleudefrance}{rgb}{0.19, 0.55, 0.91}

\usepackage{tikz}
\usepackage{pgf-pie}  
\usepackage{subfigure}

\definecolor{MayGreen}{RGB}{60, 157, 78}
\definecolor{Grape}{RGB}{112, 49, 172}
\definecolor{Ruber}{RGB}{201, 77, 109}
\definecolor{Sunray}{RGB}{228, 191, 88}
\definecolor{HanBlue}{RGB}{65, 116, 201}
\definecolor{ChineseOrange}{RGB}{246, 109, 68}
\definecolor{Rajah}{RGB}{254, 174, 101}
\definecolor{KeyLime}{RGB}{230, 246, 157}
\definecolor{LightMossGreen}{RGB}{170, 222, 167}
\definecolor{GreenSheen}{RGB}{100, 194, 166}
\definecolor{CyanCornflowerBlue}{RGB}{45, 135, 187}
\definecolor{MidnightGreen}{RGB}{0, 63, 92}
\definecolor{PurpleNavy}{RGB}{88, 80, 141}
\definecolor{Mulberry}{RGB}{188, 80, 144}
\definecolor{PastelRed}{RGB}{255, 99, 97}
\definecolor{Cheese}{RGB}{255, 166, 0}
\definecolor{ChineseOrange2}{RGB}{172, 76, 48}
\definecolor{HanBlue2}{RGB}{141,172,223}

\begin{document}

\title{Q-Turn: Changing Paradigms In Quantum Science}

\author{Ana Bel\'en Sainz}
\address{International Centre for Theory of Quantum Technologies, University of Gda\'nsk, 80-309 Gda\'nsk, Poland.}

\begin{abstract}
Quantum information is a rapidly-growing interdisciplinary field at the intersection of information science, computer science, mathematics, philosophy, and quantum science. This fruitful field of research is at the core of our developments of quantum technologies, while widening the frontiers of our fundamental knowledge, and has achieved remarkable progress in the last few decades.  
Regardless of its scientific success, quantum information is not exempt from the intrinsic features that come from the fact that scientists are  humans and members of society: both the good and the bad of our social practices leak into the scientific activity. In our scientific community, diversity and equal opportunity problems are particularly difficult to observe due to social, economic, or cultural barriers, often remaining invisible. How can our lack of awareness negatively influence the progress of science in the long term? How can our community grow into a better version of itself? 

This article reflects on how research events -- such as conferences -- can contribute to a shift in our culture. This reflection draws on what we learn from Q-Turn: an initiative triggered by postdoctoral researchers to discuss these questions, and by doing so raise awareness about diversity issues and equal opportunities in quantum science.   In addition to the high caliber of science, one of Q-turn’s main missions is to foster an inclusive community and highlight outstanding research that may be under-appreciated in other high-impact venues due to systemic biases. As well as a scientific program, Q-turn features talks and discussions on issues that affect the quantum information community, ranging from diversity and inclusion, health and mental health, to workers' rights. 

In this perspective article,  we will consider Q-Turn as an example of how a research community can work to tackle systematic biases, review the successes, and identify further points for development. 
\end{abstract}
\maketitle


\tableofcontents


\section*{The need for change}
\addcontentsline{toc}{section}{The need for change}

Picture in your mind the image of a scientist, or what is more, of a successful scientist. Ask children to do the same, and maybe make a drawing. What do we all see? In many cases, something that resembles  Emmett ``Doc'' Brown from Back to the Future. In general, we see that middle-age white man, with no real obligations or responsibilities, just their drive and mild-obsession to work on their research, day and night. Younger generations might have a different picture in mind though: that of a socially-awkward Sheldon Cooper, with no real-life struggles, whose unintentional borderline offensive behavior is perceived as funny. This perception of a scientist, built from entertainment and media exposure, is not representative of those who have the potential and interest to become scientists, however in a large proportion of cases it is not very far from the truth. 

Quantum science did not end up by accident being a field of research dominated by white-male from the global north; several factors -- that go beyond the scope of this perspective --  contributed to this. However, the narrow perspective that like-minded people bring to science may arguably hinder its progress. For the good of science -- and,  more importantly,   the well-being of the people carrying it out -- change needs to happen. 

What is it then that needs to change, and how do we make it happen? Short answer: the  basic  structure of the academic and scientific system, to account for scientists being human beings living in our far-from-perfect world. Of course, this short answer is an overly simplified and highly idealistic one. How can we do better? Spoiler alert: we do not have the full answer yet -- but we have already started taking the first steps to trigger change. 

\section{ Origins of the Q-turn workshop }

Interest in science develops within curious people since early ages. By the time we reach university, a door opens for us to a whole new world of knowledge, and it is fascinating. But as time passes, and career progresses, you start seeing beyond first impressions, and noticing your bright classmates quitting, your colleague not coming back after maternity leave, professional opportunities slipping through your hands -- a variety of small yet powerful daily facts that yell at you that you don't belong (or that if you decide to stay, that it is not going to be easy). This is the context in which many early-career researchers in quantum science find themselves, and which brought together Yelena Guryanova, Jara Juana Bermejo-Vega, and Ana Bel\'en Sainz back in the mid 2010's. 

Part of the problem is the lack of a proper safe space in which to discuss these issues. We remember once attending a faculty of physics compulsory training seminar on `implicit bias', and noticing our white male colleagues coming out of the session acknowledging the issue. So, what was different? Why did this concept that we spent so many lunch breaks trying to communicate now finally get to them? Well, this time it was presented by an expert quoting data from studies, rather by the young colleagues they meet everyday at lunch quoting instances of their personal experience. And then the idea clicked, and the seed of Q-turn started germinating: what could we achieve if we scaled this up? 

Back in 2017 the idea of Q-turn took form, and the journey of this peculiar workshop (founded by Yelena Guryanova, Jara Juana Bermejo-Vega, and Ana Bel\'en Sainz) began. Q-turn is a unique international quantum information workshop series. Its core mission is to foster an inclusive community and highlight outstanding research that may be under-appreciated in other high-impact venues due to systemic biases. Q-turn aims to facilitate a dialogue in the community over issues that affect us as a society, collectively making progress to resolve them. To this aim, Q-turn features an awareness program in addition to the focused quantum science program. Q-turn’s awareness program promotes diversity, equity, inclusion, intersectionality, responsible research, workers’ rights, as well as physical and mental health in quantum science and technology. So far there have been two editions of the Q-turn workshop, one in Florian\'opolis, Brazil, in 2018 \cite{qturn2018}, and one online edition in 2020 \cite{qturn2020}. 

 Q-turn originated as a response to a need, hence its peculiar form with strong emphasis on community issues. However, not every conference needs to have such focus in order to contribute meaningfully to a positive shift in our scientific culture. In the next section we comment on a few specific actions that conferences can endorse as best practices to help the community develop.  

\section{ Specific actions at the Q-turn workshop }

Besides its unique awareness program (details of which can be found in the next section), Q-turn takes specific actions during its organisation and implementation, which not only make the workshop a fresh gust of air for underrepresented groups in quantum science, but, hopefully, will also positively influence our quantum scientific community.  These actions are not necessarily specific to the Q-turn workshop, and hence may also be implemented at other conferences and scientific venues.   Examples of these specific actions are the following:

\begin{compactitem}
\item[]
\item \textbf{Inclusive Atmosphere}: one of the first steps is to make the members of the community feel safe and welcome. In Q-turn we make a strong conscious effort to foster an inclusive atmosphere. Regarding the building where the event is hosted, we pay special attention to choosing an accessible barrier free venue and setting up gender neutral bathrooms. In addition, we work towards securing a diverse cast of
invited speakers. For example, regarding gender diversity, in the first edition of the Q-turn workshop we had, in the scientific program, 1 man, 4 women, and 1 non-binary person, while the awareness program consisted of 4 women and 4 men. Moreover, three invited speakers were known members of the LGBTQ community. 
\item[]
\item \textbf{Code of Conduct}: we implement a code of conduct to get through personal and cultural barriers, and so set up the standard for acceptable professional conduct. This code of conduct has a clear and accessible protocol for reporting violations of it, which pays special attention towards protecting the well-being and anonymity of the involved parties. 
\item[]
\item \textbf{Diversity in Committees and Presenters}: we make a strong and conscious effort to having a diverse set of organisers, program committee members, and invited speakers, in terms of gender, ethnicity, geographical location, and area of quantum research. We devote on average more than a year to come up with suggestions for speakers and committee members, and make a careful assessment and selection. Allowing plenty of time for this is crucial for fighting unconscious biases.
\item[]
\item \textbf{Review Process for Scientific Contributions}: our review process is in constant revision to fight against systematic biases and unhealthy work practices. Our particular actions so far are: (i) assure representation of minority fields of research by having a diverse and representative program committee (PC); (ii) improve the quality of the submissions' assessments by giving each PC member a low number (about 5) of submissions to review, and by having each PC chair handle a low number of assessments -- this is achieved by recruiting a large number of PC members ($\geq 50$) and of PC chairs (4 were recruited for the 2020 edition of Q-turn); (iii) fight unconscious bias in the review process by allowing plenty of time (ideally 2 months) for the review process; (iv) fight against exploitative work relationships by discouraging the use of subreviewers. 
\item[]
\item \textbf{Mobility}:  we  allow for plenty of time (we aim for three months) between the `notification for authors' and the start of the workshop, so that presenters have time to apply for the necessary travel visas. 
\item[]
\item \textbf{Travel Grants}: we implement a \textit{travel grants programme} to cover the cost of travel, accommodation, and in some cases maintenance, for selected participants. We  target   these grants to students and young post-docs from underrepresented groups (gender, ethnicity, geographical location, area of quantum research). 
\end{compactitem}

\section{Awareness programme}

The Q-turn workshop features not only a quantum science programme, but also an `awareness' programme, the latter being the main focus of this section. In the two editions of the Q-turn workshop, there was roughly a 50/50 split in invited talks between the quantum and the awareness programmes.  Such awareness sessions are a special feature of Q-turn, but may also be implemented in other quantum scientific venues even in the form of a single invited talk or panel discussion. Organisers of established conference series might be wary of providing a space for open discussions in such topics -- and it should be done with the care it deserves so it happens in a safe space -- but we hope that, with time, speaking about community issues becomes normalised.  

\bigskip

 For context, before elaborating on the specifics of Q-turn's awareness programme, let me briefly describe its quantum programme.  Q-turn's quantum science program highlights top-quality experimental and theoretical work on quantum information technology and foundations. The fields covered include quantum foundations (causality, thermodynamics, generalised probabilistic theories), quantum communication and cryptography (algorithms, error correction, simulation), and models of quantum computation (quantum complexity theory, estimation and measurement, entanglement theory). The details of the quantum science programme go beyond the scope of this perspective. 

\bigskip

Q-turn's awareness program promotes diversity, equity, inclusion, responsible research, workers’ rights, as well as physical and mental health in quantum science
and technology. The awareness sessions (in the form of presentations or panel discussions) are run by experts on equal opportunities, inclusion and diversity, focused on the following key areas:
\begin{compactitem}
\item Representation (race; gender equality; non-binary; intersectionality; marginalised groups)
\item Conflict (unconscious bias; micro-aggressions; harassment)
\item Rights (work conditions; labour rights; contracts)
\item Health (health and mental health in academia)
\end{compactitem}

The first edition of the Q-turn featured three awareness talks plus a panel discussion, whereas its second edition featured six awareness sessions. The details of these topics, whose tins of worms we have opened, are the following: 

\begin{compactitem}
\item[]
\item \textbf{Implicit Bias}: Havi Carel (University of Bristol, UK) presented a talk on ``Implicit bias, microaggressions and chilly climates: how can we improve equality and
inclusion in academia?''. The purpose of the talk was to increase the awareness of factors that make our work environment less welcoming for some and suggest practical ways to change that. Here we were made aware of \textit{implicit bias} and \textit{stereotype threat}, as well as of the problems caused by them. We also discussed the causes of `chilly climates' within academia, caused by these factors, as well as microaggressions. 
\item[]
\item \textbf{Working in Academia}: it is usually argued that `scientists follow their passion' -- this is often used as leverage to expect people to work long hours, during weekends and holidays, for little pay (and no extra-hours remunerations). Such practices not only impact in the workers' health and well-being, but -- surprisingly for some -- also hinders productivity. Ariel Bendersky (University of Buenos Aires, Argentina) conducted an open discussion, titled ``Working in academia. The good, the bad, and the ugly'', on the toxic academic work culture, where he provided a general overview on working rights in academia. We learned about collective agreements and unions, and how false self-identification plays a role in academia: astonishingly, and sometimes unconsciously, science workers refuse to identify themselves as workers. We also discussed how `academic excellence' is a dangerous concept that enables a way to deny labour rights
\item[]
\item \textbf{Socio-economical Inclusion}: socio-economical factors play a pivotal role in implicitly post-selecting scientists that come from privileged backgrounds --  case in point,   it is not easy to progress in your academic career when you need to work two jobs and care for a struggling family. Renato Pedrosa (Unicamp, Campinas, Brazil) provided us with enlightening thoughts in his talk ``Social inclusion in higher education in Brazil: is merit and quality at peril?'', where he discussed these issues in the context of higher education in Brazil \cite{renato}. 
Since at least 2003 there had been frequent debates in Brazil about merit and its impact on the quality of education, both inside and outside of academic circles. It was then when the first public universities adopted affirmative action programmes to increase the chances of poor youngsters, including black students, of being admitted to their programs. Such affirmative actions were triggered either by the own initiative of the Universities (like University of Brasilia and
Unicamp), or mandated by the state and, later, federal law. Pedrosa presented us the data from studies on the impact of the affirmative-action policies developed on the quality of education provided by public higher education institutions (HEIs) in Brazil. 
\item[] 
\item \textbf{Black Community in Academia}: a brilliant panel of researchers opened our eyes to a myriad of perspectives that white researchers from the global north (the most-funded community quantum scientists) might have never considered. This session was led by Bárbara Rosa (Cambridge Graphene Centre, UK), Carlos Parra (Ludwig-Maximilians-University, Munich, Germany), Cornelius Mduduzi Masuku (Purdue University, US), Juan David González Calderon (Uniremington Medell\'in, Colombia), Katemari Rosa (Federal University of Bahia, Brazil), and Mathys Rennela (Leiden University, The Netherlands). Topics that we touched upon included the historical systematic bias that promotes achievements by white people, and appropriates and rebrands black and middle eastern cultural and scientific achievements as outcomes of the white community. We were also made aware of a variety of harassment behavior and micro-aggressions that black people endure on a daily basis in our academic system. 
\item[]
\item \textbf{Science is Not a Safe Space}: the scientific community is not exempt from harassing behaviours. Be it due to `cultural differences' or the toxic working culture within the hierarchical academic structure, harassment is present and experienced by many scientists ( especially   from minority groups) since early stages in their career. Harassment does not restrict to sexual harassment, but it also encompasses other types of abuse such as a supervisor over-working their students with tasks they are not meant to be doing. Ultimately, the big power imbalance and lack of awareness and acknowledgement of the problems, foster an environment where harassment leaks into the scientific work. A panel of researchers and activists led an emotionally-challenging yet eye-opening discussion on the topic of (sexual) harassment, based on studies and on personal experiences. The panel was formed by 
Emma Chapman (Imperial College London, UK), Ruth Oulton (University of Bristol, UK), and Sarah Kaiser (Q\# Community, US). 
\item[]
\item \textbf{Inclusion of People with Disabilities}: Sofia  Qvafort (University College London and Imperial College London, UK) opened our eyes to the problem of inclusion in the academic system of people with disabilities. In her talk ``Disability \& Academia'' she highlighted some of the challenges faced by disabled academics and what we can do as individuals and on a framework-level to make academia a better place for everyone. Se also shared with us her personal experience on studying and working in physics as a person with a visual impairment. 
\item[] 
\item \textbf{Mental Health in Scientific Research}: when spelled out, the low-quality of academics' mental health might not come as a surprise -- a highly-competitive time-demanding job, with poor (usually  fixed-term) work contracts, and a mobility policy that  scrapes  people from their support networks. When adding to this the personal challenges that each individual has, together with social taboos that prevent people from timely getting help, one may wonder how much progress would science experience if the main muscle of its scientific workforce was better looked after. Michelle Reynolds (University of Cambridge Staff Counselling Service, UK) and Senaida Hernández Santana (Universidad Politécnica de Madrid, Spain) walked us through an enlightening session on these topics. Interestingly, mental health struggles are more common than we think within academia \cite{crisis}, and some studies have estimated scientists' mental health quality to be on similar footing to that of healthcare practitioners \cite{rand}. 
\item[]
\item \textbf{Science Communication}: how to present your scientific findings and your visions for future research is a crucial feature of scientific activity: it impacts how you scientists and research fields are perceived by society, and vitally, how much funding is available for each field. Incidentally, intensive publicity of a particular field of topic (\textit{hype}) -- even beyond what one may realistically expect of it  --   arguably  happens, among others,  in the field of quantum technologies.  Tara Roberson (Australian Research Council Centre of Excellence for Engineered Quantum Systems, Australia) discussed the advantages and dangers of hype, how it may help to advance support for science and technology, and left us with plenty of open question on the drawbacks of hype \cite{tara}. 
\item[]
\item \textbf{Ethics in Quantum Research}: 
being driven by people, quantum research is not exempt from ethical considerations. These range across a wide spectrum, from personal scientific conduct in the workplace, to fundraising techniques and technologies' development. 
Emma McKay (McGill University, Canada) walked us through  implicit costs of scientific research that we usually don't think twice about -- how does our research impact the environment and our resources? How might our scientific activity promote or  be   sustained by colonialist practices?  In this talk, followed by an intense discussion session, we started touching the tip of the iceberg. 
\end{compactitem}

\section{Q-turn outcomes and feedback}

Q-turn 2018 hosted 114 participants, which is a great success if compared to the number of participants at similar scientific venues, such as the Conference on the Theory of Quantum Computation, Communication and Cryptography (TQC) -- 61 participants in 2014, 65 in 2015, and 104 in 2016. 
Q-turn 2020, which happened online due to the covid-19 pandemic, received over 900 registrations. We hosted over 600 participants on slack, and had 170 inspiring contributions (awareness sessions, invited and contributed talks, and posters). As a remark, we estimate around 200 of registered participants to have actually attended the Q-turn 2020 workshop in a meaningful way. 

The statistical data we computed for each edition of Q-turn was drawn from the registration forms that consented to the data being used for such purposes (714 for Q-turn 2020), as well as from the submitted satisfaction surveys (74 for Q-turn 2018 and 80 for Q-turn 2020). In this section we present some highlights of such data as well as other relevant feedback and outcomes of the Q-turn workshop. 

\bigskip

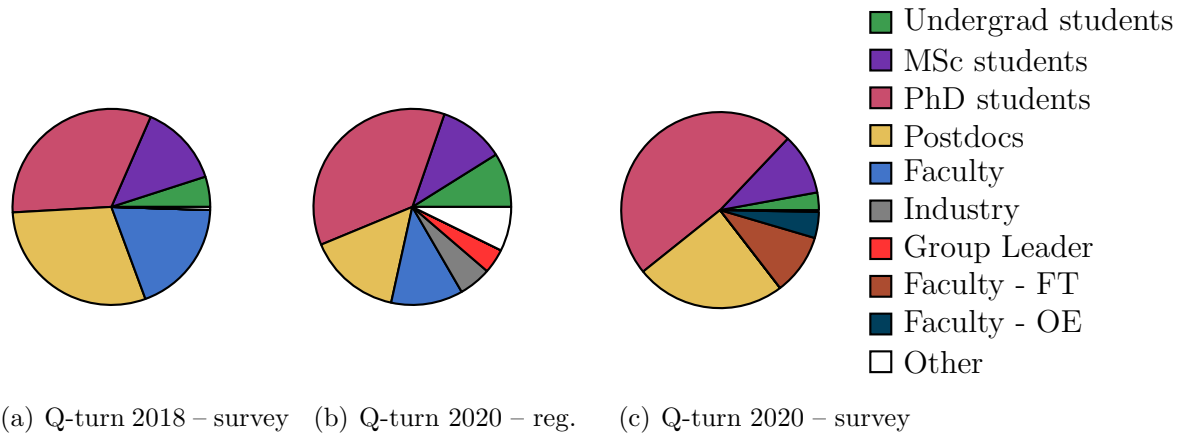
\begin{figure}
\begin{center}
\subfigure[Q-turn 2018 -- survey]{
\begin{tikzpicture}
\node at (2.2,-2.2) {};
\pie[radius=1.3,hide number,color={MayGreen, Grape, Ruber, Sunray, HanBlue, CyanCornflowerBlue, GreenSheen, KeyLime, MidnightGreen, white}]{5/,13.5/, 32.4/, 29.7/, 18.9/, 0/, 0/,  0/, 0/, 0.5/}
\end{tikzpicture}}
\subfigure[Q-turn 2020 -- reg.]{
\begin{tikzpicture}
\node at (2.2,-2.2) {};
\pie[radius=1.3,hide number,color={MayGreen, Grape, Ruber, Sunray, HanBlue, gray, red!80!white, ChineseOrange2, MidnightGreen, white}]{8.96/,10.78/, 36.55/, 15.27/, 11.76/, 5.32/, 4.06/, 0/, 0/, 7.3/}
\end{tikzpicture}}
\subfigure[Q-turn 2020 -- survey \qquad \qquad \hspace{1.5cm}]{
\begin{tikzpicture}
\pie[radius=1.3,hide number,text = legend,color={MayGreen, Grape, Ruber, Sunray, HanBlue, gray, red!80!white, ChineseOrange2, MidnightGreen, white}]{2.9/Undergrad students,10.1/MSc students, 47.8/PhD students, 24.6/Postdocs, 0/Faculty, 0/Industry, 0/Group Leader, 10.01/Faculty - FT, 4.3/Faculty - OE, 0/Other}
\end{tikzpicture}}
\end{center}
\caption{\textbf{Career stage of Q-turn participants:} (a) participants of Q-turn 2018 that filled up the satisfaction survey; (b) registered participants of Q-turn 2020; (c) participants of Q-turn 2020 that filled up the satisfaction survey. In all plots the categories for students and early career researchers are the same. In Q-turn 2018, people beyond the postdoc stage were categorised as `Faculty'. In the registration data from Q-turn 2020 people beyond the postdoc stage were instead further classified between Faculty, Industry, and Group Leader. Finally, in the survey data from Q-turn 2020 people beyond the postdoc stage were instead classified as Faculty/Group Leader on a fixed-term contract or on a permanent contract (regardless whether in academia or industry). }
\label{fig:careerstage}
\end{figure}

\noindent \textbf{Career stage profile.--} 
Already at the first edition of the Q-turn workshop we were overwhelmed by the support from the scientific community, in particular by students and early-career researchers. 
Nonetheless, participation of senior scientists at Q-turn was also satisfactorily high. As shown in Figure \ref{fig:careerstage}(a), in Q-turn 2018, 48.6\% of participants were either postdocs (29.7\%) or professor level (18.9\%) researchers, whilst 32.4\% of participants where PhD candidates and 13.5\% were master students. 
Figure \ref{fig:careerstage}(b) shows that in Q-turn 2020 the registered participants consisted of 56.26\% students (from undergraduates to PhD), 15.27\% post-doctoral researchers, and 21.14\% researchers established in academia or industry. We see, however, a different proportion of career-stages among the participants who submitted the satisfaction in Q-turn 2020, which one could argue might reflect better the statistical data of the people that meaningfully participated in the conference. This data shows a career-stage distribution of around 61\% students, 25\% postdocs, 10\% of senior scientists on non-permanent contracts, and only 4\% senior scientists in open-ended contracts, see Figure \ref{fig:careerstage}(c).  

\bigskip

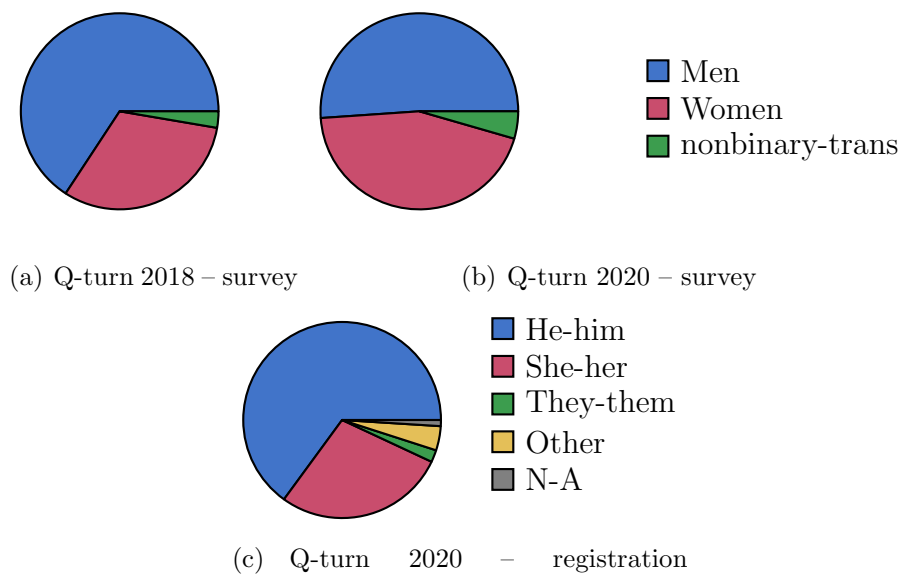
\begin{figure}
\begin{center}
\subfigure[Q-turn 2018 -- survey]{
\begin{tikzpicture}
\node at (2.2,-1.6) {};
\pie[radius=1.3,hide number, color={HanBlue, Ruber, MayGreen}]{65.8/, 31.5/, 2.7/}
\end{tikzpicture}}
\subfigure[Q-turn 2020 -- survey]{
\begin{tikzpicture}
\node at (2.2,-1.6) {};
\pie[radius=1.3,hide number,text = legend, color={HanBlue, Ruber, MayGreen}]{51.1/Men, 44.4/Women, 4.5/nonbinary-trans}
\end{tikzpicture}}
\subfigure[Q-turn 2020 -- registration \qquad \qquad \hspace{1.5cm}]{
\begin{tikzpicture}
\pie[radius=1.3,hide number,text = legend, color={HanBlue, Ruber, MayGreen, Sunray, gray}]{65/He-him, 28/She-her, 2/They-them, 4/Other, 1/N-A}
\end{tikzpicture}}
\end{center}
\caption{\textbf{Gender profile of Q-turn participants}: (a) and (b) according to the data collected by the satisfaction surveys, for Q-turn 2018 and 2020, respectively; (c) according to the data collected by the registration forms in Q-turn 2020. Even though there is still a male majority in the participants, we see how the female and gender minority representation seems higher than in other quantum venues. }\label{fig:genprof}
\end{figure}

\noindent \textbf{Gender profile.--} 
The percentage of women and gender minorities among participants of Q-turn 2018 was much higher than in typical quantum information workshops: 31.5\% were women and 2.7\% were non-binary or transgender, while 65.8\% were men (see Fig.~\ref{fig:genprof}(a)). 
Remarkably, a very similar ratio was observed among talk presenters, with 31.8\% women and 4.5\% transgender or non-binary, while 63.8\% were men. 
Many participants, organizers, and program committee members appraised the high ratio of women and gender minorities -- one of the greatest achievements of the conference, in particular, since it \textit{did not have any gender quota}

Q-turn 2020 assessed the gender profile of participants by their preferred pronouns. We see that the percentage of participants that use the pronouns ``he/him'' roughly coincides with the percentage of men attending Q-turn 2018. This is quite remarkable, given that the geographical profile of the conference is quite different in the two editions, and that the number of responses analysed in Q-turn 2020 is about an order of magnitude higher than in Q-turn 2018. This seems to suggest that these numbers might represent the community overall, or at least the overall quantum community curious about community building. Figure \ref{fig:genprof}(c) presents a graphical depiction of the collected data. 

Finally, there is the data collected through the satisfaction survey of Q-turn 2020. Of the received answers, 51.1\% were by men, 44.4\% were by women, and 4.5\% where by non-binary/trans participants (see Fig.~\ref{fig:genprof}(b)). Hence the proportion of men that replied to the satisfaction survey was quite lower than the proportion of women and gender minority participants that did so. 

\bigskip

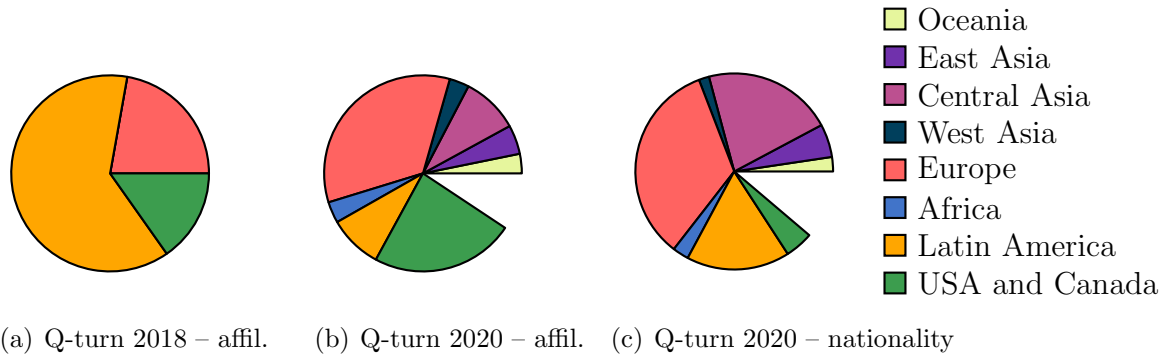
\begin{figure}
\begin{center}
\subfigure[Q-turn 2018 -- affil.]{
\begin{tikzpicture}
\node at (2.2,-1.6) {};
\pie[radius=1.3,hide number,color={KeyLime, Grape, Mulberry, MidnightGreen, PastelRed, HanBlue, Cheese, MayGreen}]{0/, 0/, 0/, 0/,22.22/,0/,62.5/, 15.28/}
\end{tikzpicture}}
\subfigure[Q-turn 2020 -- affil.]{
\begin{tikzpicture}
\node at (2.2,-1.6) {};
\pie[radius=1.3,hide number,color={KeyLime, Grape, Mulberry, MidnightGreen, PastelRed, HanBlue, Cheese, MayGreen}]{3.22/, 4.8/, 9.38/, 3.22/,34.16/,3.5/,8.82/,23.52/}
\end{tikzpicture}}
\subfigure[Q-turn 2020 -- nationality \qquad \qquad \hspace{1.5cm}]{
\begin{tikzpicture}
\pie[radius=1.3,hide number,text = legend, color={KeyLime, Grape, Mulberry, MidnightGreen, PastelRed, HanBlue, Cheese, MayGreen}]{2.38/Oceania, 5.46/East Asia, 21.29/Central Asia, 1.68/West Asia, 33.74/Europe, 2.66/Africa, 16.9/Latin America, 4.62/USA and Canada}
\end{tikzpicture}}
\end{center}
\caption{\textbf{Geographical profile of Q-turn participants}: (a) and (b) according to their country of affiliation, for Q-turn 2018 and 2020, respectively; (c) according to their nationality for Q-turn 2020. We see that the in-person 2018 edition that happened in Brazil had a strong participation of people from the Americas, whilst lacking participants from Asia, Oceania, and Africa. Q-turn 2020, in turn, given its virtual character and focus on geographical diversity achieved a more varied attendance. One curiosity to notice in Q-turn 2020 is the large fraction of attendees that work in the USA\&Canada, despite the low number of people originally from those countries.  } \label{fig:geoprof}
\end{figure}

\noindent \textbf{Geographical profile.--} 
Q-turn 2018 assessed the geographical profile of its participants in terms of their country of affiliation (see Fig.~\ref{fig:geoprof}(a)). 
The strategic location of Q-turn in Brazil enabled a significant participation from the Americas. 
The biggest representation per country was Brazil,  showing  that  Q-turn 2018 succeeded  in  engaging  with  the  local  scientific community.
Scientists from European institutions also attended the workshop. 
However, we see a hole in participation from Asia, Oceania, and Africa. We believe this to be a combination of how long/expensive the trip from those places to Brazil is, as well as the high need for travel support that some institutions from countries in those continents have. 

Q-turn 2020 took the covid-19 pandemic as an opportunity to work on geographic inclusion, and the conference was organised so that it could be enjoyed by people all across the globe: it hosted more than 50 hours of event across three time zones. Remarkably, participants joined from institutions from about 50 countries.
Q-turn 2020 hence broadened the scope of the statistical analysis to two geographical aspects of the registered participants: their country of affiliation and their country of nationality. 
Regarding country of affiliation, 34.16\% of the participants came from Europe, 23.52\% from the USA and Canada, 9.38\% from Central Asia, 8.82\% from Latin America, 4.8\% from East Asia, 3.5\% from Africa, 3.22\% from West Asia, and 3.22\% from Oceania (see Fig.~\ref{fig:geoprof}(b)). We see that this geographical distribution is quite different from that of Q-turn 2018; I believe this to be due to the virtual format of the workshop, as well as the lack of necessity for travel support schemes. 
Regarding nationality of the participants, 33.74\% are from Europe, 21.29\% are from Central Asia, 19.6\% are from Latin America, 5.46\% are from East Asia, 4.62\% are from the USA and Canada, 2.66\% are from Africa, 2.38\% are from Oceania, and 1.68\% are from West Asia (see Fig.~\ref{fig:geoprof}(c)). The countries where most of our participants came from are India, Brazil, and England, respectively. We also observe some amusing features, such as $\sim 24$\% of the participants come from the USA and Canada, whilst only $\sim 5$\% are nationals of those countries (and may even work in a different one).

\bigskip

\noindent \textbf{Satisfaction surveys summary.--} 
In general the participants were very positive about both the quantum and the awareness programmes. They found the opportunity to discuss community issues very valuable, and they overall feel empowered to have an active role on triggering change. 

A recurrent comment is the need to have more time, dedicated spaces, and blackboards to further discuss science and `awareness' topics. 
Having a less-packed schedule and dedicated spaces may be a way to tackle this. The need for a `quiet room' where to unwind and work without disturbance was also frequently mentioned. 
The participants also highlighted the importance and need for a `transparent and well-funded' financial support scheme.

Regarding the awareness programme, participants of Q-turn 2018 mentioned the need to discuss `mental health in academia' as well as the `eurocentric' character of academic research and `journal publication policies'. In addition, participants from Q-turn 2020 remarked that the majority of the awareness speakers come from a white or global-north background, and that more diversity there would be appreciated. 

Q-turn 2020, being an only-online conference, presented participants with a unique (at the time) set of aspects to assess. Most welcomed the virtual character of the conference as more inclusive and environmentally-friendly, and suggest that Q-turn continues in a hybrid form. In particular, participants were happy that the event was organised across three different time zones, which they claim improved accessibility to it. One drawback of the virtual character of Q-turn 2020 was that the participants struggled to socialise as they would in in-person conferences, and suggest that some strategies be devise to promote virtual interaction (e.g., using platforms such as gather-town).

\section{Looking forward: challenges and opportunities}

The Q-turn workshop has opened the door to a unique opportunity for improving science: it provides the space where to make people aware of the issues that hinder our scientific activity, and where to leverage bright people (our quantum scientists) to brainstorm solutions. Q-turn also provides a unique space where to scientifically interact with other scientific communities within your own research field, which has the potential to considerably increase the quality and impact of the scientific activity.
Nonetheless, there are many challenges towards making the best use of the doors that the Q-turn workshop opens, and here I discuss a few that we have strongly encountered in these past five years.  Some of these challenges are specific to a relatively young conference series, although others may be experienced at quantum  scientific events in general. 

\begin{compactitem}
\item[] 
\item \textbf{Our Own Biases}: how to make Q-turn grow beyond the close environment of the original Q-turn players? For Q-turn 2020 we implemented an anonymous way for people to suggest invited speakers and topics for both the quantum and the awareness programmes. This was a good first step, but we did not receive as many suggestions as we had hoped. We need the active participation of the broad quantum community to make a fair and representative selection in the long run. 
\item[]
\item \textbf{Participation}: we have substantial support and participation from young members of the community, but this means we are also lacking the expertise and influence of established scientists, professors, and group leaders -- those that have the best footing to \textit{take risks} and \textit{implement change}. Support from such players could also make Q-turn an attractive venue for people who are not yet convinced of Q-turns scientific quality, probably due to systematic and unconscious biases. How to engage these parts of the community is still an open question, and at the moment we have tackled the active and worldwide promotion of our activities. We set up a mailing list (currently with ~220 members) to keep interested scientists and Q-turn enthusiasts up to date on the activities related to Q-turn. Our events are also advertised in international mailing lists and on posters and notice boards in universities and institutes around the world. The Q-turn workshop is also promoted during talks at outreach events and conferences.
\item[]
\item \textbf{Funding}: a main aim of the Q-turn workshop is to bring down the barriers set up by privilege: it is crucial that our workshop is attended by minority groups, which in  many   cases do not have access to funds for conference attendance. Hence, Q-turn has aimed to provide financial support for participants, and minimise conference fees. This requires an immeasurable amount of fundraising work, which has not proven as successful  as we had hoped.  A main barrier here is the lack of `prestige' in such a young conference series, especially given the early-career stages of the founding members and conference organisers. We need funding bodies to see beyond their usual checkpoints and start to substantially support activities like ours. 
\item[]
\item \textbf{Reluctance to Change}: this is the hardest challenge that we face. It may not come as a surprise that not everyone is enthusiastic about the activities of Q-turn. To name one,  it seems  we will need years  before   the existence of codes of conduct  becomes commonplace and cease to be a contentious issue.  We need people to not take our topics personally, and keep an open mind towards the needs of the always-changing scientific community. Equal opportunities initiatives are not meant to threaten or belittle privileged groups, but rather to raise awareness of, and trigger action to overcome, the issues that  we face  as a community -- and we need the support and involvement of our more privileged colleagues to make the community a better place. 
\end{compactitem}

\section*{Conclusions}
\addcontentsline{toc}{section}{Conclusions}

 The Q-turn workshop's journey has taught us many lessons, and not only for the Q-turn initiative: lessons for research events in quantum science, and lessons for the quantum science community itself. 

Scientific events in general play a key role in shaping the community and the way we do science, and so have a golden opportunity to help trigger change. As I have discussed in this article, there are many specific actions that scientific conferences and events other than Q-turn can also implement; actions that do not incur extra costs but that can already help, such as most of the points in Section~2. In addition, scientific events can consider generating a space for raising awareness and discussing community issues, such as it has been done at the `Bristol Quantum Information Technologies' workshop series since 2019 and the `Quantum Correlations, Contextuality and All That Again$^n$' workshop series since 2017. 

From a community point of view, a comforting lesson to learn is that `we are not alone in this'. The proportion of scientists that are keen for change and in need of it is more than we think and more than we can see. As a community, we need to become agents of our scientific practices, to give change a chance to happen. The Q-turn workshop is a clear example of that. 

\bigskip 

Throughout our Q-turn journey, we have sparked the flame of awareness and change. We have only touched upon the tip of the iceberg, and there is still a lot  more  to learn and discuss. More importantly, we are yet to see the impact of the Q-turn workshop on concrete actions aimed at implementing positive change. For this, the involvement of and support from higher-up players is vital. It is, however, reassuring and hopeful that there is a seed in the community that is motivated to holistically improve quantum science, and moreover, that there is a place -- Q-turn -- for them to engineer change from.

\section*{Acknowledgments}
I am extremely grateful to John H.~Selby for his feedback on this manuscript and his constant support during my Q-turn journey. I am also grateful to Yelena and Juani for all their work and support, especially when we started walking our Q-turn dream. We are grateful to the Foundational Questions Institute (FQXi) for their generous financial support of the two editions of the Q-turn workshop.
ABS acknowledges support by the Foundation for Polish Science (IRAP project, ICTQT, contract no.2018/MAB/5, co-financed by EU within Smart Growth Operational Programme). 
All pie charts were drawn using Tikz.

\end{document}